\documentclass[twocolumn,prl,aps,superscriptaddress]{revtex4-2}
\usepackage{color}
\usepackage{graphicx}
\usepackage{natbib}
\usepackage{subfigure}

\newcommand{\nc}{\newcommand}
\nc{\on}{\operatorname}
\nc{\wt}{\widetilde}
\nc{\Wick}{{\mathbb :}}
\nc{\R}{{\mathbb R}}

\newcommand{\beq}{\begin{equation}}
\newcommand{\eeq}{\end{equation}}
\newcommand{\bmul}{\begin{multline}}
\newcommand{\emul}{{\end{multline}}}
\newcommand\beqa{\begin{eqnarray}}
\newcommand\eeqa{\end{eqnarray}}
\newcommand\bea{\begin{array}}
\newcommand\eea{\end{array}}
\newcommand\ba{\begin{array}}
\newcommand\ea{\end{array}}

\newcommand{\neqa}{\nonumber\end{eqnarray}}

\nc{\CH}{{\mathcal H}}
\nc{\Db}{{\bar D}}
\nc\comment[1]{}

\nc{\CM}{{\mathcal M}}
\nc{\CN}{{\mathcal N}}

\newcommand{\re}{\relax{\rm I\kern-.18em R}}

\nc{\meV}{{\mathrm{\,meV}}}
\nc{\cG}{{\mathcal G}}

\renewcommand{\)}{\right)}

\comment{ \)}
\renewcommand{\bar}{\overline}

\nc{\al}{{\alpha}}

\begin{document}

\title{Fate of density waves in the presence of a higher order van Hove singularity}
\author{Alkistis Zervou}
\affiliation{Department of Physics and Centre for the Science of Materials, Loughborough University, Loughborough LE11 3TU, UK.\\}
\author{Dmitry V. Efremov}
\affiliation{Department of Physics and Centre for the Science of Materials, Loughborough University, Loughborough LE11 3TU, UK.\\}
\affiliation{Leibniz Institute for Solid State and Materials Research Dresden, Helmholtzstrasse 20, D-01069 Dresden, Germany.\\}
\author{Joseph J. Betouras}
\affiliation{Department of Physics and Centre for the Science of Materials, Loughborough University, Loughborough LE11 3TU, UK.\\}

\date{\today}

\begin{abstract}
Topological transitions in electronic band structures, resulting in van Hove singularities in the density of states, can considerably affect various types of orderings in quantum materials. Regular topological transitions (of neck formation or collapse) lead to a logarithmic divergence of the electronic density of states  (DOS) as a function of energy in two-dimensions. In addition to the regular van Hove singularities, there are higher order van Hove singularities (HOVHS) with a power-law divergences in DOS. By employing renormalization group (RG) techniques,  we study the fate of a spin-density wave phase formed by nested parts of the Fermi surface, when a HOVHS appears in parallel. We find that the phase formation can be boosted by the presence of the singularity, with the critical temperature increasing by orders of magnitude. We discuss possible applications of our findings to a range of quantum materials such as Sr$_3$Ru$_2$O$_7$, Sr$_2$RuO$_4$ and transition metal dichalcogenides.

\end{abstract}
\maketitle

\paragraph{Introduction.} Phase transitions, due to spontaneous symmetry breaking, with the emergence of an order parameter, are closely connected to specific features of the electronic band structure of itinerant systems. For example,  various density waves appear in systems with nesting in the electronic band structure, i.e. the spectrum of the electronic excitations close to the Fermi level is characterized by $\varepsilon(\mathbf{p+Q})\approx - \varepsilon(\mathbf{p})$ where the vector $\mathbf{Q}$ is the nesting vector.  Well known representatives of density waves are the archetypal chromium \cite{Fawcett}, cuprates \cite{Wise}, iron pnictides \cite{Mazin,Dai}, organics \cite{Monceau}, and transition metal dichalcogenides \cite{Ausloos}. 
Intriguingly,  in a range of these materials the band structure hosts energetically close-by singularities in the density of states $\nu$ (DOS), which have been conjectured often to be crucial ingredients stabilizing the emergent phases \cite{vanHove, Kim, Kang, Sherkunov-Betouras, Classen_2020}.

Singularities and the associated divergence of DOS are a signature of the Fermi surface's topological transitions~\cite{Shtyk, Efremov-Betouras}.
The two more well-known cases dealt by Lifshitz \cite{Lifshitz} in his original work were the appearance or collapsing of a neck and the appearance or collapsing of a pocket in Fermi surface. 
The former case was the ordinary van Hove singularity (VHS), with the Fermi surface locally consisting of a pair of intersecting straight lines. These two types of Fermi surface topological transitions have been observed along with their non-trivial consequences due to interactions in a wide range of quantum materials including cuprates, iron arsenic and ferromagnetic superconductors, cobaltates, Sr$_2$RuO$_4$, heavy fermions \cite{Liu, Coldea, Okamoto, Yelland, Khan, Bernhabib, Slizovskiy-Chubukov-Betouras, Aoki, Sherkunov-Chubukov-Betouras, Barber,Stewart, Markiewicz, Baumberger, Hicks, Hicks2}.

However, HOVHS display more exotic Fermi surface topological changes that lead to even more intriguing properties. They have been associated with exotic phenomena such the non-trivial magnetic and thermodynamic properties in $\mathrm{Sr_3 Ru_2 O_7}$~\cite{Efremov-Betouras}, correlated electron phenomena in twisted bilayer graphene near half filling~\cite{Yuan}, the so called supermetal with diverging susceptibilities in the absence of long range order~\cite{Isobe1} and unusual Landau level structure in gated bilayer graphene \cite{Shtyk}.  Recently, a classification scheme for Fermi surface topological transitions and their associated DOS divergence was developed \cite{Chandrasekaran-Shtyk-Betouras-Chamon, LiangFu}  as well as a method to detect and analyze them  \cite{Chandrasekaran-Betouras-2}, while  the effects of disorder were also studied \cite{Chandrasekaran-Betouras}.

Here, we study the general question about the fate of a density wave, spin-density wave (SDW) or current charge-density wave (CDW), that is formed due to nesting of two parts of the Fermi surface when the Fermi energy is tuned so that a Fermi surface topological transition with HOVHS in the DOS at nearly the Fermi level emerges. If the degree of nesting is not significantly changed due to the HOVHS, the density wave phase, as naively expected, can be suppressed. Surprisingly, we find that it can get boosted, depending on the strength of the bare couplings in the Hamiltonian.

\paragraph{Model.}  We take three patches within the first Brillouin zone (BZ). Two of them (patch 1 and 2), with DOS $\nu_0$ per spin, are nested both in the presence and absence of patch 3, which is the one associated with the singular DOS. The dispersion relations are $\varepsilon_1 ({\bf k} )= - \varepsilon_2 ({\bf k} + {\bf Q}) = v_F (k_x - k_F)$, were $v_F$ is the Fermi velocity and $k_F$ is the Fermi momentum of the two nested patches. The dispersion relation of the third patch with respect to the chemical potential $\mu$ is modelled by  $\varepsilon_3({\bf k}) = \alpha k^2 + \gamma (k_x^4+k_y^4-6 k_x^2k_y^2) - \mu$.
In the present work we consider the problem at the quantum critical point assuming $\alpha =0$ for simplicity. This is the form that has been recently considered for a higher order VHS in $\mathrm{Sr_3 Ru_2 O_7}$~\cite{Efremov-Betouras}.  The resulting DOS per spin for patch 3 is then:
\begin{equation}
\nu(\varepsilon)= A_4 {|\varepsilon|}^{-1/2}. \label{eq:Density-2}
\end{equation}
where $A_4=\alpha_4/\sqrt{\gamma}$ with
$\alpha_4=\frac{1}{16}\frac{1}{\pi^{3/2}}\frac{\Gamma\left(\frac{1}{4}\right)}{\Gamma\left(\frac{3}{4}\right)} \approx 0.033$,  $\Gamma$ is the gamma-function and $\gamma$ is measured in units of $1/\nu_0$.
Below, we take $1/\gamma = 100 \nu_0$. We consider all possible short-range electron-electron interactions  
%
allowed by symmetry and obeying the conservation of momenta.  We assume ${\bf Q}$ to be incommensurate, as such Umklapp processes are not relevant.  Taking into account all possible {\it relevant} two-particle-interactions involving fermions in the three patches, the effective Hamiltonian reads:
%

\begin{widetext}

\begin{eqnarray}
\nonumber
    & \mathcal{H}&= \int d{\bf k} \sum_{\sigma=\uparrow, \downarrow} \sum_{a=1,2,3} \varepsilon_a ({\bf k}) c^\dagger _{a \sigma} ({\bf k}) c_{a\sigma} ({\bf k}) 
    + \bar{g}_1 \int \left\{d{\bf k}_i\right\}   \sum_{\sigma \sigma'}  c^\dagger _{1\sigma} ({\bf k}_1) c^\dagger _{2\sigma'} ({\bf k}_2) c_{2\sigma'} ({\bf k}_3) c_{1\sigma} ({\bf k}_4)  \\
    \nonumber
     &+& \bar{g}_2 \int \left\{d{\bf k}_i\right\}   \sum_{\sigma \sigma'}   c^\dagger _{1\sigma} ({\bf k}_1) c^\dagger _{2\sigma'} ({\bf k}_2) c_{1\sigma'} ({\bf k}_3) c_{2 \sigma} ({\bf k}_4)
      + \bar{g}_3 \int \left\{d{\bf k}_i\right\}   \sum_{\sigma \sigma'}   c^\dagger _{1\sigma} ({\bf k}_1) c^\dagger _{2\sigma'} ({\bf k}_2) c_{3\sigma'} ({\bf k}_3) c_{3 \sigma} ({\bf k}_4) \\
      \nonumber
      &+ & \bar{g}_4 \int \left\{d{\bf k}_i\right\}   \sum_{\sigma \sigma'}   c^\dagger _{3\sigma} ({\bf k}_1) c^\dagger _{3\sigma'} ({\bf k}_2) c_{3\sigma'} ({\bf k}_3) c_{3 \sigma} ({\bf k}_4)  
       + \bar{g}_5 \int \left\{d{\bf k}_i\right\}   \sum_{\sigma \sigma', {a=1,2}}    c^\dagger _{a \sigma} ({\bf k}_1) c^\dagger _{3\sigma'} ({\bf k}_2) c_{3\sigma'} ({\bf k}_3) c_{a \sigma} ({\bf k}_4) \\
       &+& \bar{g}_6 \int \left\{d{\bf k}_i\right\}   \sum_{\sigma \sigma', {a=1,2}}    c^\dagger _{3\sigma} ({\bf k}_1) c^\dagger _{a \sigma'} ({\bf k}_2) c_{3\sigma'} ({\bf k}_3) c_{a \sigma} ({\bf k}_4)  + h.c.
\end{eqnarray}

\end{widetext}
\noindent where $a$ labels the patches, $\sigma, \sigma'$ are spin indices, patches 1 and 2 are the nested ones and are taken equivalent. The $\bar{g}_1$ term describes density-density  interactions between patches 1 and 2, $\bar{g}_2$ takes into account exchange interactions between patches 1 and 2,  $\bar{g}_3$ describes pair transfer between patch 3 and patches 1,2, while $\bar{g}_4$ describes density-density within patch 3 and $\bar{g}_5$ and $\bar{g}_6$ density-density and exchange interactions respectively between patch 3 and each of patches 1 and 2 (Fig. 1). The interactions which are solely within patch 1 or patch 2 are irrelevant and are not presented in the Hamiltonian as their particle-particle and particle-hole bubbles associated with them are small and can be neglected in parquet Renormalization group (pRG).  The conservation of momentum is assumed. 
In principle the effective Hamiltonian and the relative strength of the interactions can be obtained starting with a short-range interaction (Hubard and Hund's model), when the orbitals that play dominant role at each patch are known. We leave the parameters quite general to account for different possibilities. In the following, we use  dimensionless $g_i$'s with $g_i = \nu_0 \bar{g}_i$.

 \begin{figure}[t!]
    \begin{center}
   \includegraphics[scale=0.40]{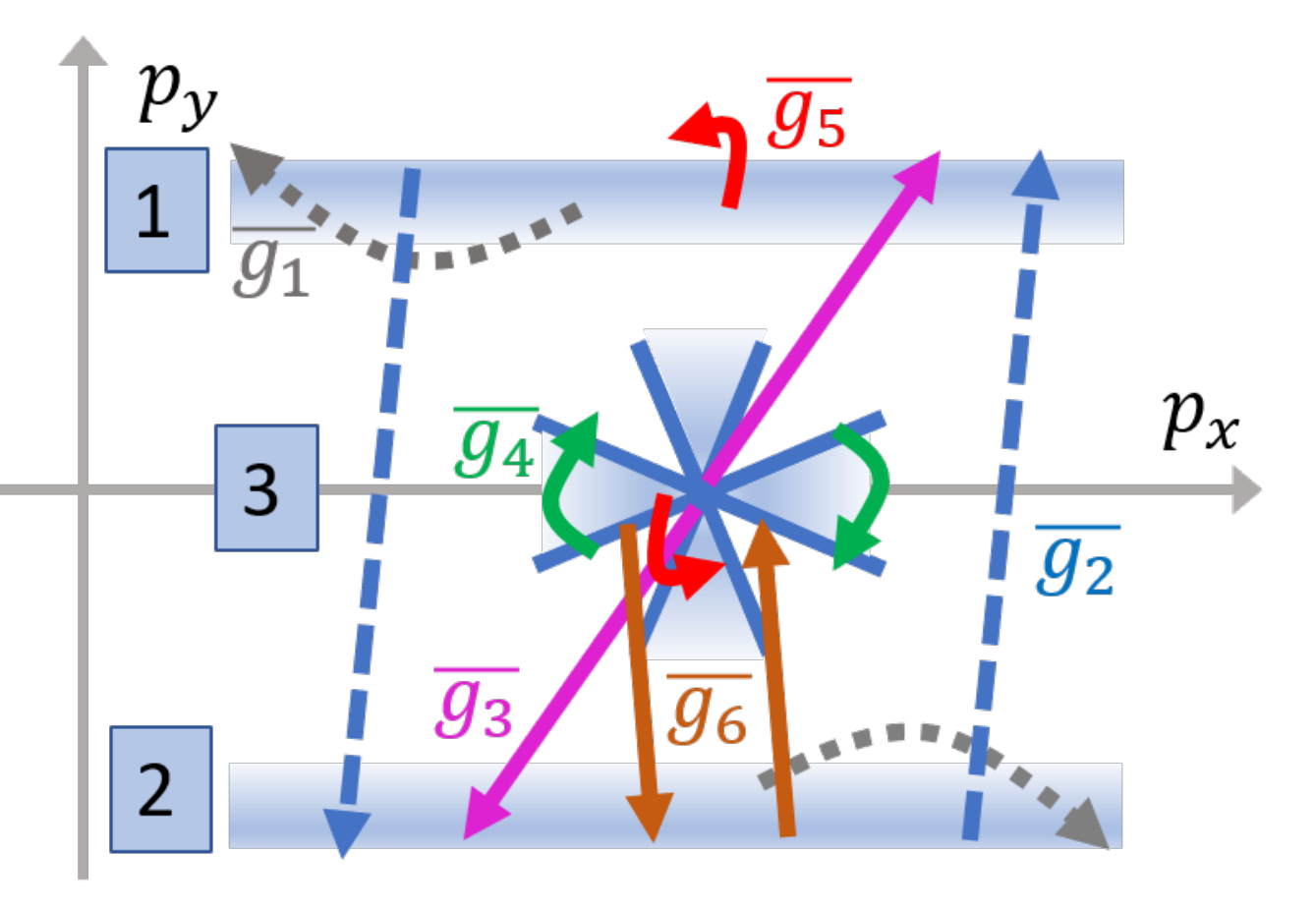}
     \end{center}
 \caption{Schematically, the interactions of the Hamiltonian Eq.(2).}   
       \label{fig:Fig1}
\end{figure}

\paragraph{Particle-particle and particle-hole bubbles.}
As the geometry of the system dictates there are two characteristic momenta. The first one is the nesting vectors {\bf Q}, which connects patches 1 with 2. 
The second one ${\bf \tilde{Q}}$
connects patches 1 with 3 and 2 with 3. In the following, we denote by $\Pi$ the particle-hole and by $C$ the particle-particle non-interacting susceptibilities respectively. The particle-hole susceptibility  for the patches 1 and 2 is $\Pi_{ph} ^{(12)} (\omega=0, {\bf Q})  = T \sum_n \int d{\bf k} G_1(\omega_n,{\bf k}) G_2( \omega_n,{\bf Q}+ {\bf k})$,
where $G_{1,2}(\omega_n, {\bf k}) =\left[i\omega_n -\varepsilon_{1,2}({\bf k})\right]^{-1}$ are the corresponding Green's functions. Similarly, for the particle-particle bubble $C_{pp} ^{(12)} (\omega=0, {\bf q}=0) = 
T \sum_n \int d{\bf k} G_1(\omega_n,{\bf k}) G_2( -\omega_n,{\bf -k}) $. 
In the RG process, the energy integration over the regions $[\varepsilon-\delta \varepsilon, \varepsilon]$ and $[-\varepsilon, -\varepsilon+\delta \varepsilon]$ results in: 
\begin{equation}
	 -\delta\Pi^{(12)}(\varepsilon, {\bf Q}) = \delta C^{(12)}(\varepsilon, q = 0) =  \nu_0  \frac{ \tanh \left( \frac{\varepsilon}{2T}\right)}{\varepsilon} \delta \varepsilon.
	 \label{eq.delta.pi12}
\end{equation}
%

For patch 3 the leading divergences of free-particle susceptibilities are associated with the particle-particle $C^{(33)}$ and particle-hole $\Pi^{(33)}$ bubbles at zero momentum transfer.  More specifically the energy integration over the regions $[\varepsilon-\delta \varepsilon, \varepsilon]$ and $[-\varepsilon, -\varepsilon+\delta \varepsilon]$ leads to \cite{SM}:
\begin{eqnarray}
 \delta \Pi^{(33)}\!\! \left(\varepsilon,q = 0\right) &\!\!=\!\! 
 & -\frac{\nu\left(\varepsilon\right)}{2T} \cosh^{-2} \frac{\varepsilon}{2T} \delta \varepsilon,
\label{eq:delta_Pi_33}
 \\ 
 \delta C^{(33)}\!\! \left(\varepsilon,q=0\right)&\!\!=\!\!
&\nu\left(\varepsilon\right)\frac{\tanh\left(\frac{\varepsilon}{2T}\right)}{\varepsilon} \delta \varepsilon.
\label{eq:delta_C_33}
 \end{eqnarray}

%
\noindent where $\nu(\varepsilon)$ is given by Eq. (\ref{eq:Density-2}). Comparing Eqs. (\ref{eq.delta.pi12}-\ref{eq:delta_C_33}) we see three different energy dependencies of the susceptibilities. At large energies $\varepsilon\gg T$ the slope of the susceptibilities in particle-particle and particle-hole channels for patches 1,2 are slowly decaying functions of  $\varepsilon$ as $-\delta \Pi^{(12)}/\delta \varepsilon =\delta C^{(12)}/\delta \varepsilon \sim 1/\varepsilon$, which leads to the standard ultraviolet $\log(\Lambda)$ and infrared $\log(\varepsilon)$ divergences.  In contrast, the susceptibilities for the patch 3 decay much faster with $\varepsilon$  as  $\delta C^{(33)}/\delta \varepsilon \sim \varepsilon^{-3/2}$ and $\delta\Pi^{(33)}/\delta \varepsilon \sim -\exp (-\varepsilon/T)/ (T \sqrt{\varepsilon})$ when $\varepsilon \geq T$. If $\varepsilon < T$ then $\delta C^{(33)}/\delta \varepsilon$ and $\delta\Pi^{(33)}/\delta \varepsilon$ scale as $1/ (T \sqrt{\varepsilon})$ with opposite sign.
The divergence at lower limit is removed by the temperature factor in the case of the particle-particle channels and it is completely absent in the case of the particle-hole channel.   We have also checked the contribution of the susceptibilities $C^{(13)}({\bf Q}-{\bf \tilde{Q}}), \Pi^{(13)}({\bf Q}-{\bf \tilde{Q}})$ (similarly $C^{(23)}$ and $\Pi^{(23)}$) and found them to be negligible in comparison to $\Pi^{(12)},C^{(12)}, \Pi^{(33)}, C^{(33)}$. Therefore, in the next we do not include them. This simplification allows us to work in energy space.

{\it pRG equations.} 
We employ one loop pRG, which can be connected to functional RG  (e.g.  Ref.[\onlinecite{Classen_2020}]) and has been used successfully e.g. Refs.\cite{Eremin, Nandkishore1, Chubukov-Maiti, Sherkunov-Betouras, Nandkishore2}. However, a cutoff scheme is needed for the implementation of the procedure.
In this work we use energy shell RG and the cutoff scheme is:
\begin{eqnarray}
\nonumber
\frac{1}{i\omega_{n}-\varepsilon\left(k\right)}\rightarrow\frac{\Theta\left(\left|\varepsilon\left(k\right)\right|-\Lambda\right)}{i\omega_{n}-\varepsilon\left(k\right)}
\end{eqnarray}
which interpolates between the zero propagator and the bare propagator
as the cutoff $\Lambda$ changes between $0$ and $\infty$, whereby one obtains the energy shell pRG. 

For the RG equations we only include the terms with the most diverging susceptibilities and redefine the RG flow parameter $ L= \log\left(\frac{\Omega}{\Lambda} \right)$, where $\Omega = 1/\nu_0$ is the bandwidth of the pockets 1 and 2. 
The resulting set of the differential equations has the following form: 
%
%
%
%
%
\begin{eqnarray}
\nonumber
&\dot{ g_1}&  =  \eta_1  g_1 ^2    
- \eta_2 g_3^2  
+ 2 \eta_3 g_5 ( g_6 -  g_5)  \\
\nonumber
&\dot{g_2}& =  2\eta_1 (g_1 g_2- g_2^2) 
- \eta_2 g_3^2  + \eta_3 g_6^2
      \\
&\dot{ g_3}& =  - \eta_2 g_{3}g_{4}  
\label{eq.RGeq}
\\
\nonumber
& \dot{g_4}& = \left( \eta_3 - \eta_2  \right) g_4^2  \\
\nonumber
&\dot{g_5}& = \eta_3 g_4 (g_6 - g_5) \\
&\dot{g_6}& = \eta_3 g_6 g_4 \nonumber
\end{eqnarray}
with 
\begin{eqnarray}
\eta_1 &=& \tanh\left(\frac{ e^{-L}}{2\nu_0T}\right)
\\
\eta_2 &=& \frac{\alpha_4}{(\nu_0 \gamma)^{1/2}}e^{L/2}  \tanh \left(\frac{ e^{-L}}{2\nu_0 T}\right)
\label{eq.etas}
\\ 
\eta_3 &=& \frac{\alpha_4}{(\nu_0 \gamma)^{1/2}} e^{-L/2} \frac{1}{2 (\nu_0 T)  \cosh^2 \left( \frac{ e^{-L}}{2\nu_0T} \right)}
\end{eqnarray}
The functions $\eta_{1,2,3}$ determine the low-energy cut-offs at T, for all bubbles  $\Pi^{(12)}$, $C^{(12)}$,  $\Pi^{(33)}$ and $C^{(33)}$.
{\it pRG analysis.}  To understand the effect of the singularity, we compare the flow of $g_1$ and $g_2$ which are responsible for the formation of DWs, without and in the presence of patch 3. 
First we consider the patches 1 and 2 without 3. 
Considering for simplicity T $\to 0$, the expression $\tanh\left(\frac{\epsilon}{2T}\right) \to 1$ and:
	\begin{eqnarray} 
			\dot{g}_1 &=& g_1^2, \nonumber\\ 
			\dot{g}_2 &=& 2(g_1 g_2 - g_2^2) \label{eq.flow.eq.only1and2}
	\end{eqnarray}
The solution of system Eq. (10): 
\begin{eqnarray} \label{eq.flow.eq.only1and2}
		g_1 &=& \frac{g_1^0}{1-g_1^0L}, \nonumber \\ 
		g_2 &=& \frac{1}{2}\frac{g_1^0}{1-g_1^0L} + \frac{1}{2} \frac{u^0}{1+u^0 L} .
\end{eqnarray}
where $u^0=2 g_2^0- g_1^0$ and $g_{1,2}^0$ are the bare values (initial conditions) of $g_{1,2}$.
Following Ref.\cite{Eremin}, the SDW and CDW vertices read:
\begin{eqnarray}
\nonumber
	\Gamma_{SDW} &=& g_1(L) = \frac{g_1^0}{1-g_1^0L} \\
	\Gamma_{CDW} &=& g_1 - 2 g_2 = -\frac{u^0}{1+u^0 L}
\end{eqnarray}  
The equations of the gap functions then are: 
\begin{eqnarray}
	\frac{d \Delta_{\lambda}}{d L} = \Gamma_{\lambda} \Delta_{\lambda} 
\end{eqnarray}  
where $\lambda$ = SDW or CDW, with solutions in the absence of patch 3:
\begin{eqnarray}
\nonumber
	\Delta_{SDW} = \frac{\Delta_{SDW}^0}{1-g_1^0L} \mbox{~~and~~} \Delta_{CDW} = \frac{\Delta_{CDW}^{0}}{1+u^0 L}
\end{eqnarray}
The important point is that there are two independent channels  leading to SDW and CDW order parameters. SDW is formed when the critical $L \to 1/g_1^0$ while CDW when  $L \to -1/u^0$. The presence of patch 3 renormalizes strongly these critical values at finite temperature.

Returning to the full set of the flow equations Eqs. (6) to investigate the effect of the patch 3, we calculate the critical temperature in the presence and absence of  patch 3. The critical temperature is defined as the temperature of the divergence of $\Delta_{CDW}$ or $\Delta_{SDW}$.   We fix the bare $g_1^0=0.07$ and $g_2^0=0.03$ and vary the remaining bare coupling constants. The results are presented at Figs. 2 and 3. The critical temperature in the absence of patch 3 is T$_c \approx 7 \times 10^{-7}$ (the flow of $g_1$, $g_2$ and the vertices are presented in the SM).
When the patch 3 is present, then the rest of $g_i$'s come into play. For the same initial values as before for $g_1$ and $g_2$ and for the same set of initial values of $g_4=0.1$, $g_5=0.1$, $g_6=0.17$, we present the behavior of the vertices for two cases $g_3^0=0.03$ (Fig. 2) and $g_3^0=0.15$ (Fig. 3). In Fig. 2 the T$_c$ of SDW is boosted by 2 orders of magnitude to T$_c = 4.1 \times10^{-5}$ while in Fig. 3, the larger $g_3^0$ promotes the formation of CDW, with T$_c$ again two orders of magnitude greater than the $T_c$ of SDW without patch 3. As it is evident, the presence of the singularity, with nonzero $g_i$'s for $i=3, 4, 5, 6$, renormalises strongly $g_1$ and $g_2$. 
Depending on the initial conditions the effect of the singular patch 3 on the DWs is summarised by the following statements: for small values, as physically expected, of the pair-transfer $g_3^0$ (i) the SDW is destroyed when $g_5^0 > g_6^0$ (ii) if $g_5^0 < g_6^0$ it is very much enhanced (with T$_c$ enhanced by potentially orders of magnitude). Otherwise, larger values of $g_3^0$ promote the formation of current CDW. Full investigation of parameters, in momentum space, will be presented elsewhere. 

\begin{figure}[t!]
    \begin{center}
    \includegraphics[scale=0.3]{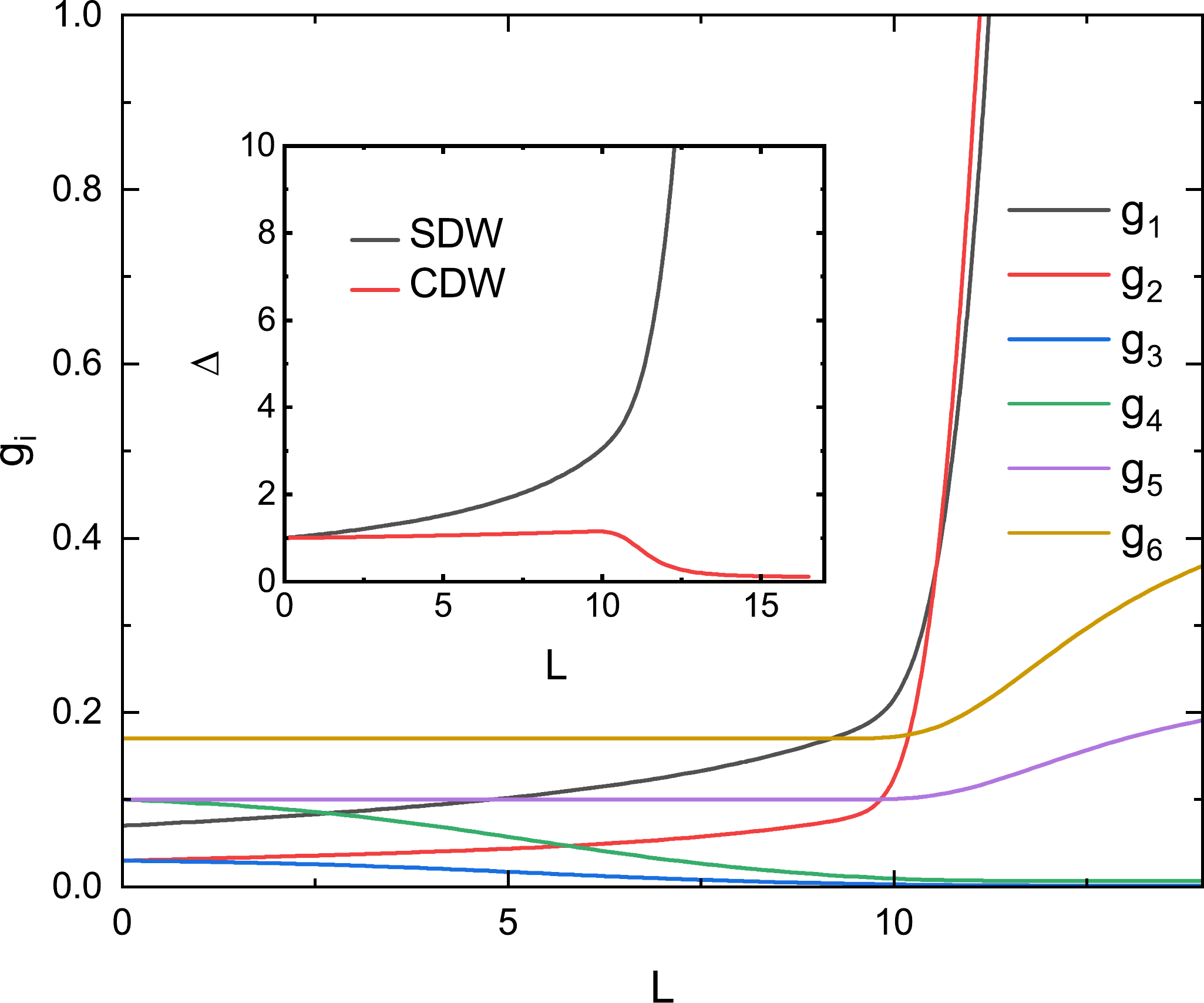}
  \end{center}
 \caption{The flow of $g_i$'s in the presence of patch 3, for initial values  $g_1^0=0.07$, $g_2^0=0.03$, $g_3^0=0.03$, $g_4^0=0.1$, $g_5^0=0.1$ and $g_6^0=0.17$ at T=T$_c$. The critical temperature for the SDW formation is found now at T$_c \propto 10^{-5}$, two order of magnitude higher than in the absence of patch 3 where T$_c \propto 10^{-7}$. In the inset, $\Delta$ denotes $\Delta_{SDW}$ or $\Delta_{CDW}$ as calculated from Eq. (12).}   
       \label{fig:Fig2_patch3_SDW}
\end{figure}

\begin{figure}[t!]
    \begin{center}
    \includegraphics[scale=0.3]{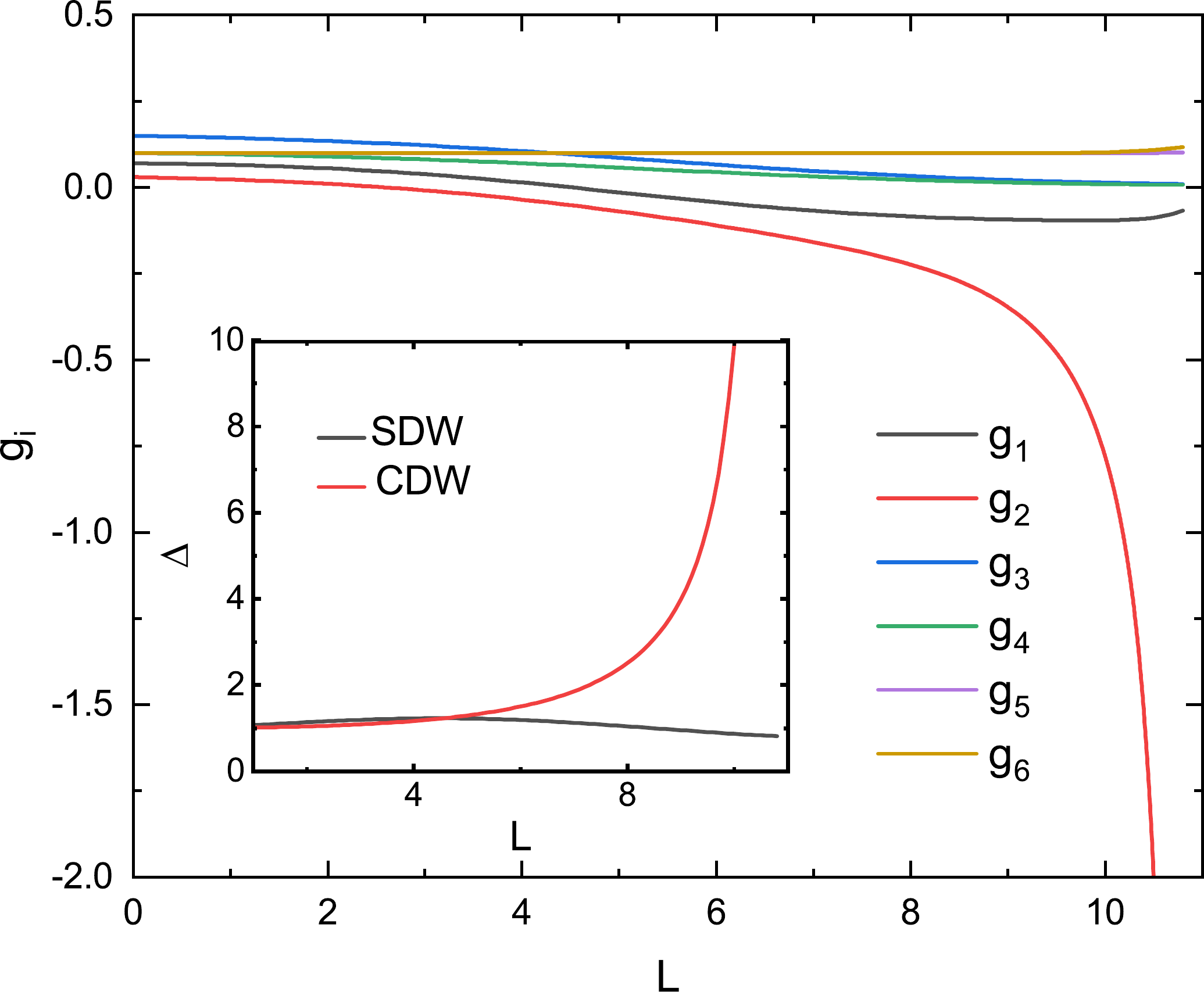}
  \end{center}
 \caption{The flow of $g_i$'s in the presence of patch 3, is for the same initial values as in Fig. 2 for all $g_i^0$ except from $g_3^0$ which is now $g_3^0=0.15$ at T=T$_c$.
 The critical temperature now corresponds to the transitions to CDW and is T$_c \propto10^{-5}$.}   
       \label{fig:Fig3_patch3_CDW}
\end{figure}

In the absence of patch 3, the flow equations Eq. (6) are such that the repulsive interactions cannot be reverted to attractive. It is easy to see that in this case $g_1$ can only grow, while $g_2$ cannot change the sign.  
The presence of patch 3 makes possible for $g_1$ and $g_2$ to change sign and become negative (attractive) \cite{SM}, due to overscreening effect caused by the HOVHS \cite{Eremin, Chubukov-Maiti,Chubukov-Maiti2}. This is a very interesting feature of the model. For lower temperatures we searched for the possibility $g_1$ and/or $g_2$ to diverge (signature of superconducting instability) but we concluded that, at low temperature, this is not possible for this model.

{\it Discussion.} In this study we have considered the effect of a HOVHS on the formation of a density wave (in particular SDW) due to nesting of other parts of the FS. The scattering through the patch with the singular DOS can have very important consequences, depending on the bare values of the interactions. It can definitely destroy the phase but surprisingly it can also amplify the formation of the density wave and increase the corresponding T$_c$ by orders of magnitude. The boosting of the DW formation happens as long as the exchange interactions between patch 3 and each of the nested patches is greater than the corresponding density-density interactions. In the opposite case, the SDW gets destroyed. 
We also find that if the initial value of the pair-transfer between patch 3 and the other two patches is strong enough, a SDW can be turned to a current CDW.  The different bare values of the interaction mimics material-specific effects such as the specific geometry in the BZ of the patches and the orbitals involved which can be different in nature. One major question is the feasibility of stronger exchange interactions in comparison to density-density ones. This is possible in multi-orbital systems \cite{comment, Isobe2}.

%

Recently, many surprising experimental results of Sr$_3$Ru$_2$O$_7$ \cite{Rost} were explained assuming the presence of a HOVHS in a magnetic field  \cite{Efremov-Betouras}. The reason that SDW phases (A and B) \cite{Lester} only appear adjacent to a HOVHS by tuning the external magnetic field, although the same nesting vector connect the edges of the $\gamma$-bands as well as other parts of the FS which respond less drastically to the magnetic field, was not explained. 
Although the difference of the present general theory to the experiments on  Sr$_3$Ru$_2$O$_7$ is that the latter is a case of SU(2) symmetry breaking as HOVS appears when the minority spins in the $\gamma$-bands sink below Fermi energy, the present work can explain in principle why the SDW was detected only when the HOVHS appeared. Indeed, there are parts of the FS that can provide the nesting which are almost insensitive to the applied magnetic field while the $\gamma$-bands are strongly affected. These bands are responsible for the formation of the HOVHS which, in turn, can boost the formation of the SDW to a critical temperature that is measurable. Therefore,  the mechanism presented here can be the key one to explain the SDW formation through the effect of the singularity at the centre of the $\gamma$-bands to the other nested pieces of the BZ. Also importantly,  the existence of an SDW in Sr$_2$RuO$_4$ when it is strained and transverses a VHS has been established \cite{Grinenko, Li}. Our work could be relevant to this finding but further work is needed.

Our theory could also explain, in principle, the CDW formation in 1T-VSe$_2$ \cite{Yang, Trott} where a VHS is present \cite{Feng} and a HOVHS at the $\Gamma$ point of the BZ is  seen in DFT calculations \cite{Esters}. This geometry corresponds to a dispersion relation with a term $\propto k^6 \cos(6 \phi)$ that can boost the formation of the CDW.  The difference is that the initial setup of the problem should favor a CDW instead of a SDW formation. We expect this theory to apply to many different materials in similar situation.

\begin{acknowledgments}
We thank  Anirudh Chandrasekaran, Claudio Chamon, Clifford Hicks and Ioannis Rousochatzakis for helpful discussions and, especially, Andrey Chubukov and Frank Kruger for a critical reading of the manuscript. We acknowledge the contribution of Garry Goldstein in the early stages of the work. The work was supported by the EPSRC grants No. EP/P002811/1 and EP/T034351/1 (JJB). D.E. acknowledges partial financial support from DFG through the project 449494427 and  Volkswagen Foundation through ``Synthesis, theoretical examination and experimental investigation of emergent iron-based superconductors". 
\end{acknowledgments}


\begin{widetext}

{\center{$\;\;\;\;\;\;\;\;\;\;\;\;\;\;\;\;\;\;\;\;\;\;\;\;\;\;\;\;\;\;\;\;\;\;\;\;\;\;\;\;\;\;\;\;\;\;\;\;\;\;\;\;\;\;\;\;\;\;\;$SUPPLEMENTAL MATERIAL}}

\setcounter{equation}{0}
\setcounter{figure}{0}
\setcounter{table}{0}
\setcounter{page}{6}
\renewcommand{\theequation}{S\arabic{equation}}
\renewcommand{\thefigure}{S\arabic{figure}}
\renewcommand{\bibnumfmt}[1]{[S#1]}
\renewcommand{\citenumfont}[1]{S#1}


\section{Calculation of ${\bf \Pi}^{(33)}$}
\label{appendix1}

The derivation for the polarization bubble $\Pi^{(33)} ({\bf q} = 0, T)$ is provided here:


\begin{equation} 
    \Pi^{(33)} (\textbf{q} \approx 0, T) = \int\int \frac{d^2p}{(2 \pi)^2}  \frac{n(\varepsilon_3(\textbf{p})) -n(\varepsilon_3(\textbf{p+q}))}{\varepsilon_3 (\textbf{p}) -\varepsilon_3 (\textbf{p+q})} = \int\int\frac{d^2p}{(2 \pi)^2}  \frac{\partial{n(\varepsilon_3(\textbf{p}))}}{\partial{\varepsilon_3 (\textbf{p)}}} 
    \end{equation}
    
\noindent where $n(\varepsilon_3(\textbf{p}))$ and $\varepsilon_3 (\textbf{p})$ are the Fermi-Dirac distribution and the dispersion relation of patch 3 respectively. Using the hyperbolic coordinates: $\eta = p^4 \sin4\phi$ and $\xi = p^4 \cos4\phi$ and taking into account the Jacobian of the transformation, the integral is transformed to:
  
  \begin{eqnarray}  
 \Pi^{(33)} (\textbf{q} = 0, T) &=&  \int\int\frac{1}{(2\pi)^2}d\xi d\eta  \quad \frac{1}{4}\frac{1}{(\xi^2+\eta^2)^{\frac{3}{4}}} \frac{\partial n(\xi)}{\partial \xi}\nonumber\\
&=&  -\frac{1}{(2\pi)^2}\frac{1}{16T}\int_{-\infty}^{+\infty}\int_{-\infty}^{+\infty} d\xi d\eta  \quad \frac{1}{(\xi^2+\eta^2)^{\frac{3}{4}}}  \cosh^{-2}\left(\frac{\xi}{2T}\right)  \nonumber\\
   &=& -\frac{1}{(2\pi)^2}\frac{1}{16T}2\int_{0}^{+\infty} d\xi  \quad \frac{\sqrt{\pi} \Gamma \left (\frac{1}{4} \right) }{\sqrt{\xi} \Gamma \left (\frac{3}{4}\right)} \cosh^{-2}\left(\frac{\xi}{2T}\right) \nonumber\\
    &=& - \frac{1}{2T}  \int_{0}^{+\infty} d\xi  \;\; \nu(\xi) \cosh^{-2}\left(\frac{\xi}{2T}\right)
 \end{eqnarray}

\noindent where $\nu(\xi)$ is the density of states for patch 3 (Eq.(1) of the main text for both spin species as we consider SU(2) symmetric case). We have also used the following relation: 
\begin{equation}
\int_{-\infty}^{+\infty} d\eta \frac{1}{(\xi^2+\eta^2)^{\frac{n-1}{n}}}=\frac{\sqrt{\pi} \Gamma \left (\frac{1}{2} -\frac{1}{n}\right) }{\xi^{\frac{n-2}{n}} \Gamma \left (1 -\frac{1}{n}\right)}, \qquad \xi>0.
\end{equation}

From Eq. (S2),  we obtain Eq. (4) of the main text.

Performing the same change of variables to hyperbolic coordinates and following the same steps for the particle-particle channel, we get to Eq. (5) of the main text for $\delta C^{(33)}$.


\newpage

\section{SDW and CDW vertices}
\label{appendix2}

Here we provide the equation for the vertices. Diagrammatically it is:

\begin{figure}[h]
\centering
\includegraphics[width=0.45\textwidth]{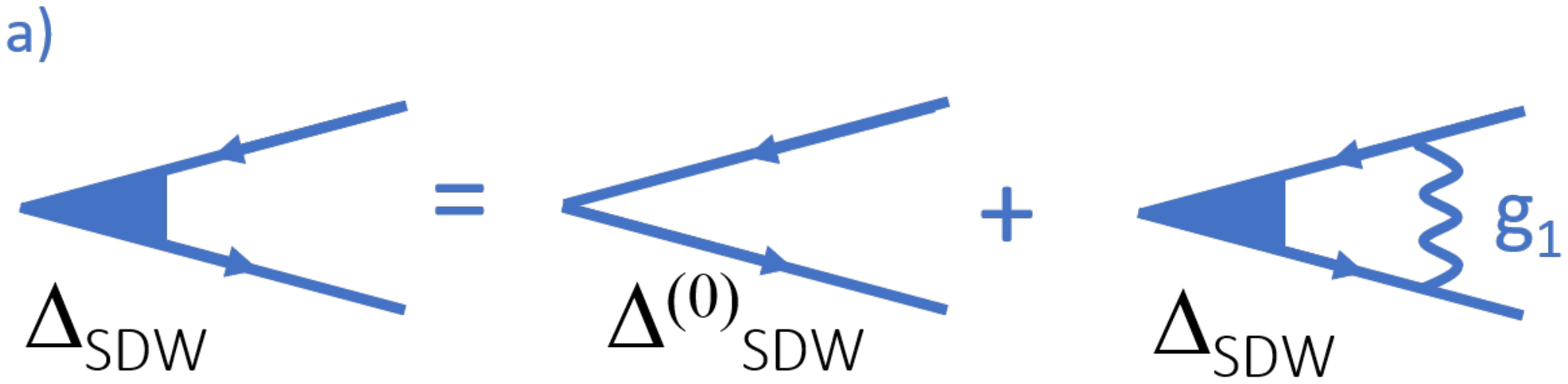} 
\includegraphics[width=0.45\textwidth]{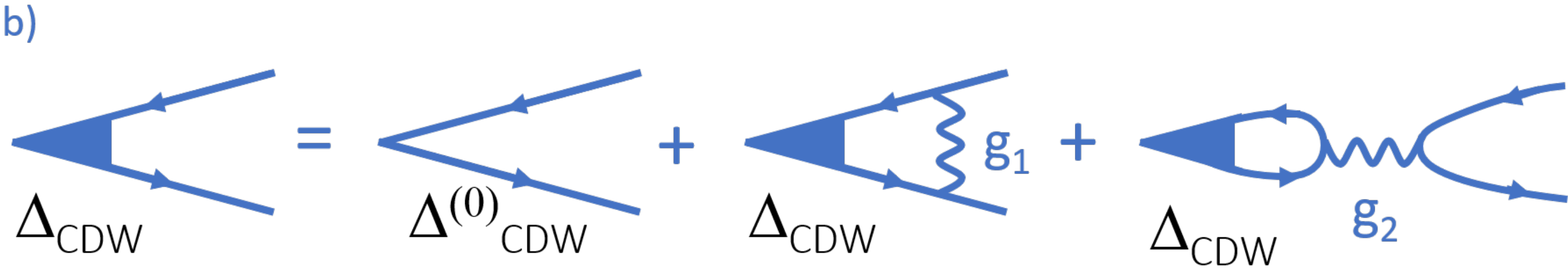} 
\caption{\label{fig:Graphene_R_Hall} The diagrammatic equations for the order parameter of  SDW and CDW and the associated vertices. (a) for the SDW and (b) for the CDW. }
\end{figure}

For the CDW case and for the vertex corrections related to process $g_2$ there is a factor (-2), due to the associated internal bubble and the summation over the internal spin degree of freedom. As the result, as written in the main text for the vertices: $\Gamma_{SDW}\left(L\right)=g_{1}\left(L\right)$ and $\Gamma_{CDW}\left(L\right)  = g_{1}\left(L\right)-2g_{2}\left(L\right)$.

\section{Flow of  $g_1$ and $g_2$ and vertices in the absence of patch 3.}
\label{appendix3}

 \begin{figure}[h]
    \begin{center}
   \includegraphics[scale=0.25]{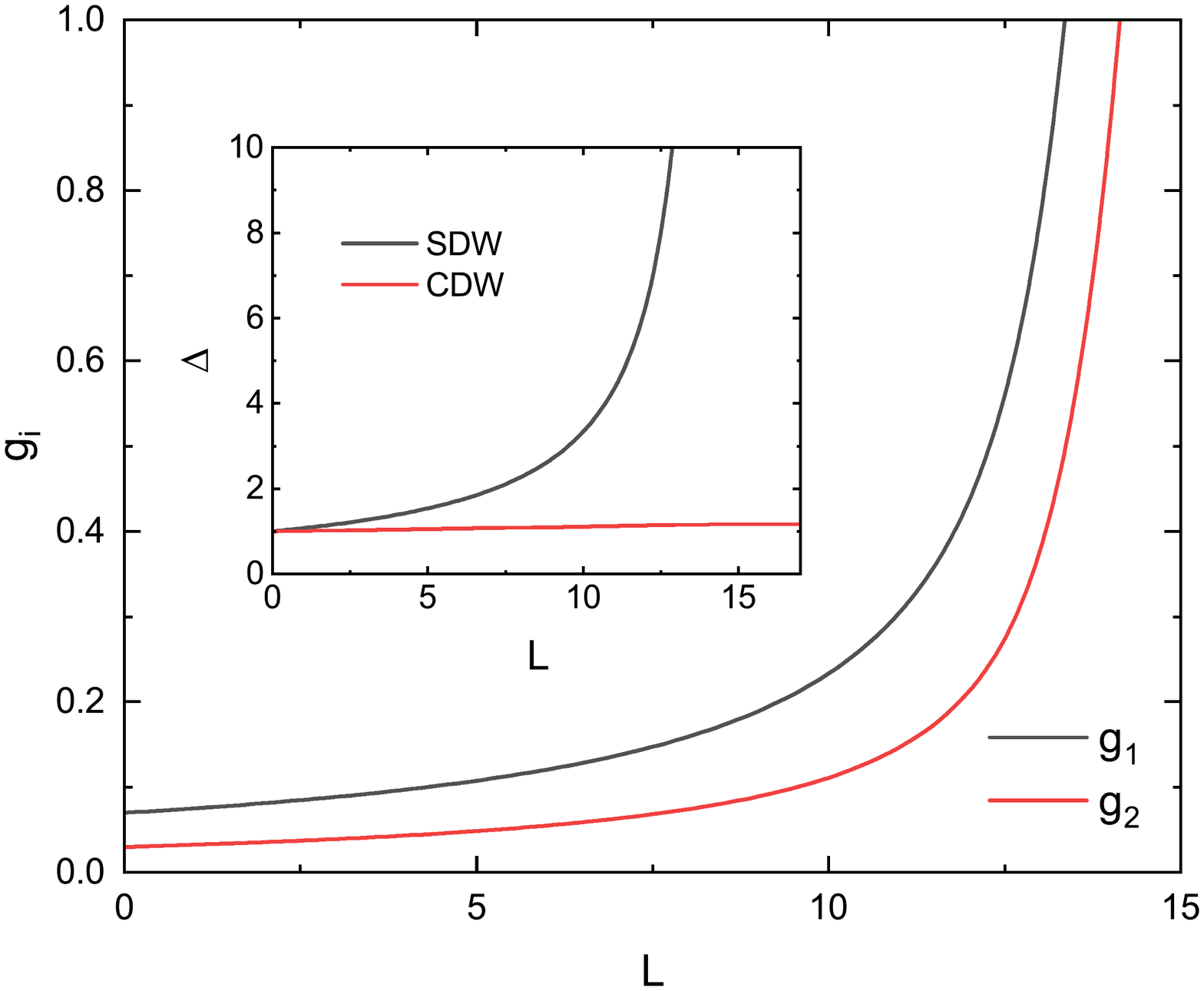}
     \end{center}
 \caption{The flow of $g_1$ and $g_2$ in the absence of patch 3, for initial values $g_1^0 = 0.07$ and $g_2^0=0.03$ at T=T$_c$. The critical temperature for the SDW formation has a $T_c \simeq 7 \times10^{-7}$. $\Delta$ denotes $\Delta_{SDW}$ or $\Delta_{CDW}$ as calculated from Eqs. (13).}   
       \label{fig:Fig1_no_patch3}
\end{figure}

For completeness, we present the results in the case where only $g_1$ and $g_2$ are non-zero (absence of patch 3). The relevant system from Eqs (6) is solved at finite temperature and the critical temperature is calculated from Eqs. (13). This critical temperature is compared with the case which patch 3 is present in the main text to establish that patch 3 may act as an amplifier to the SDW formation. 

\section{More results from RG calculations for {\bf $1/\gamma=100 \nu_0$} for $g_1^0 < g_2^0$.}
\label{appendix3}

We present some more numerical results, for the case $g_2^0 > g_1^0$ which support the conclusions of the main text.
%
\begin{figure}[htp]
\centering
\includegraphics[width=0.3\textwidth]{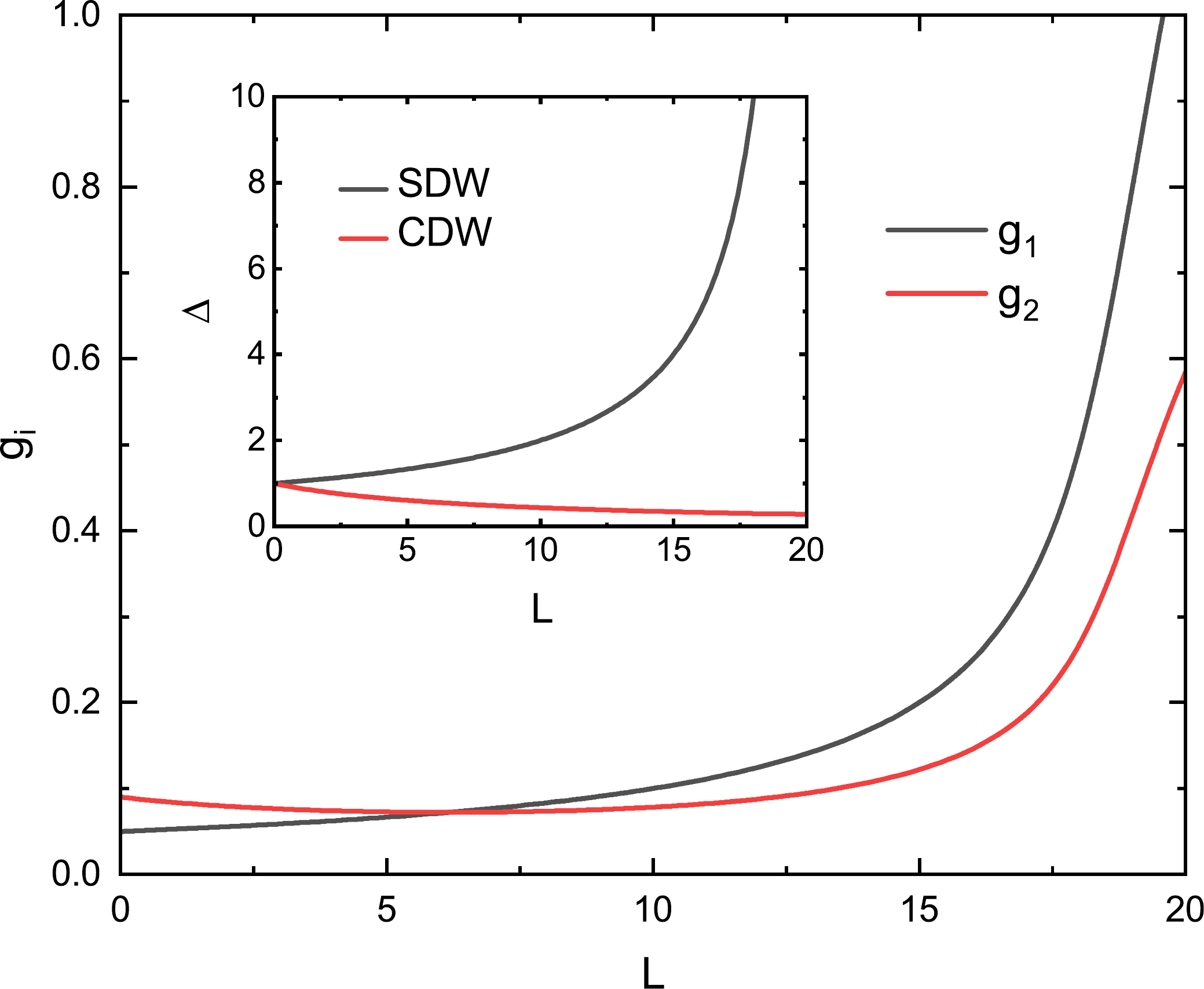} 
\includegraphics[width=0.3\textwidth]{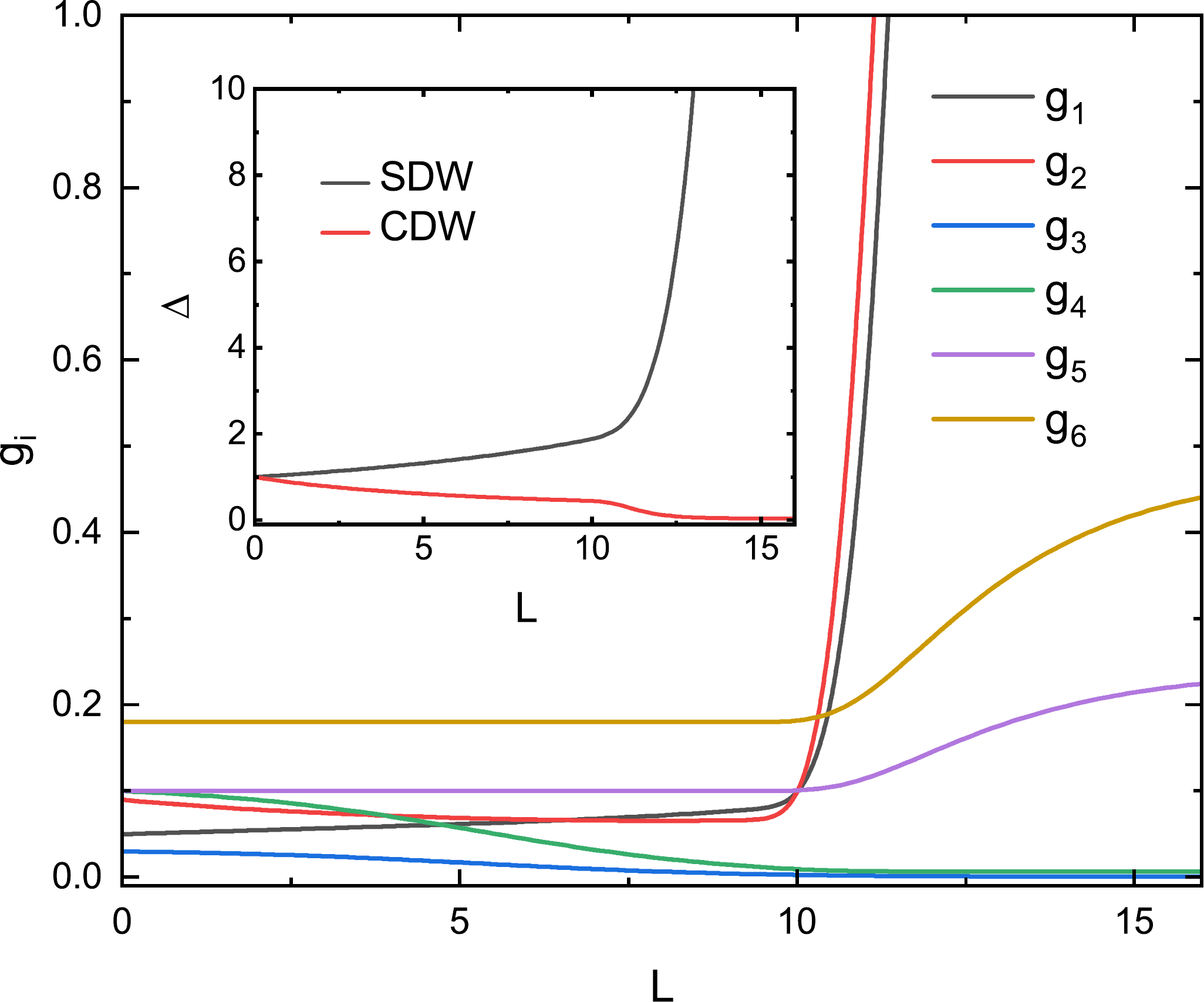} 
\includegraphics[width=0.3\textwidth]{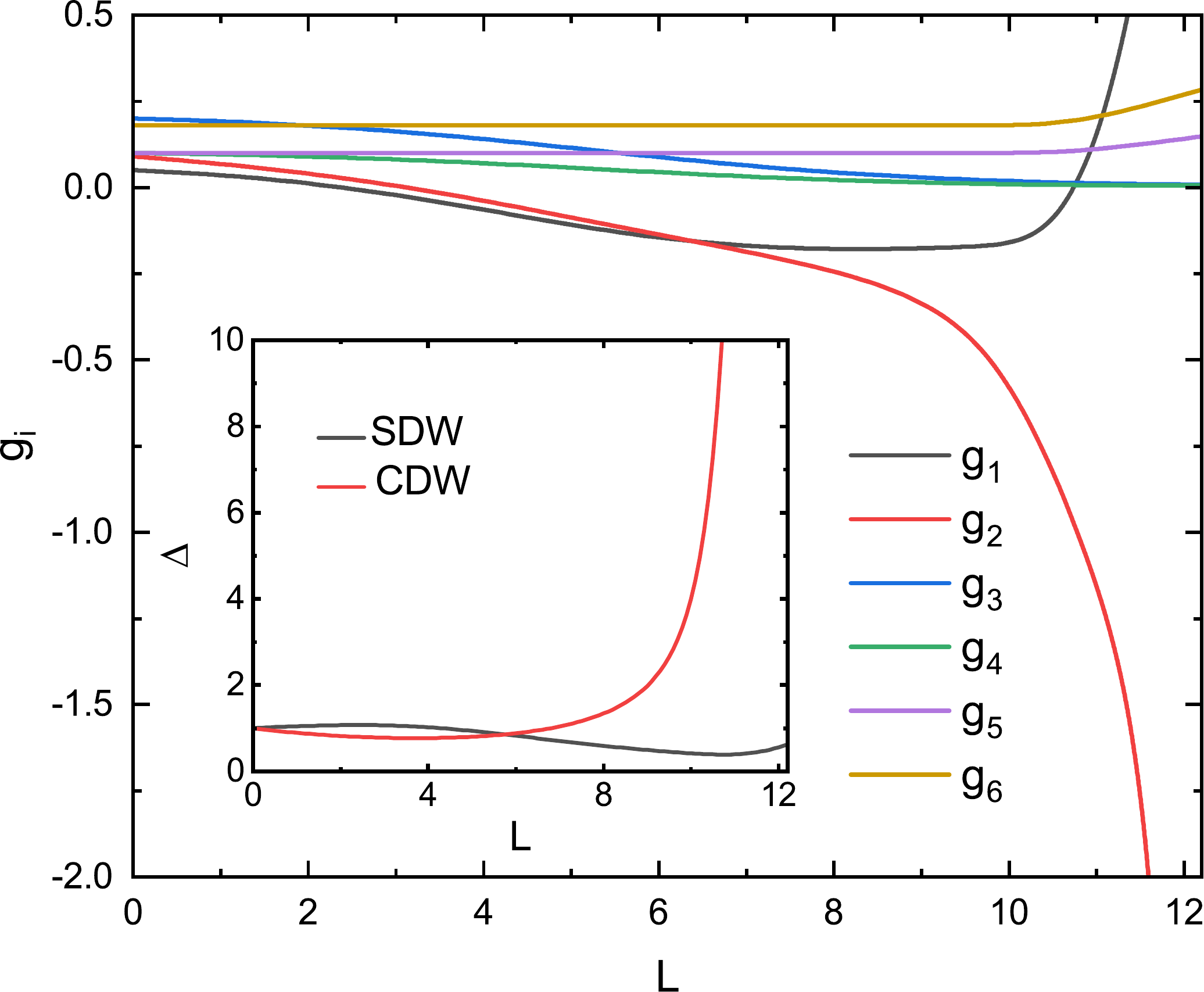} 
\caption{The initial conditions are $g_1^0=0.05$, $g_2^0=0.9$. Then in the absence of patch 3, T$_c= 5 \times 10^{-9}$ (left panel). In the presence of patch 3, there is strong enhancement of the SDW T$_c= 10^{-5}$  for $g_3^0=0.03$, $g_4^0 = 0.1$,  $g_5^0=0.1$, $g_6^0=0.18$ (middle panel). Then for right one: with $g_3^0=0.2$ and and all the rest of the initial conditions the same, the phase is CDW with T$_c= 9 \times 10^{-6}$. As we see, the enhancement of T$_c$ is more than three orders of magnitude in both cases.}
\end{figure}

\section{Results for  {\bf $1/\gamma = 10 \nu_0$}}
\label{appendix4}
Here we present results for $1/\gamma=10 \nu_0$ and the same values of $g_i$'s as in the main text for comparison. The conclusions are the same. There is strong enhancement of the values of T$_c$ although less strong than the case $1/\gamma = 100 \nu_0$ due to the prefactor in the DOS which is proportional to 1/$\sqrt{\gamma}$.

\begin{figure}[htp]
\centering
\includegraphics[width=0.35\textwidth]{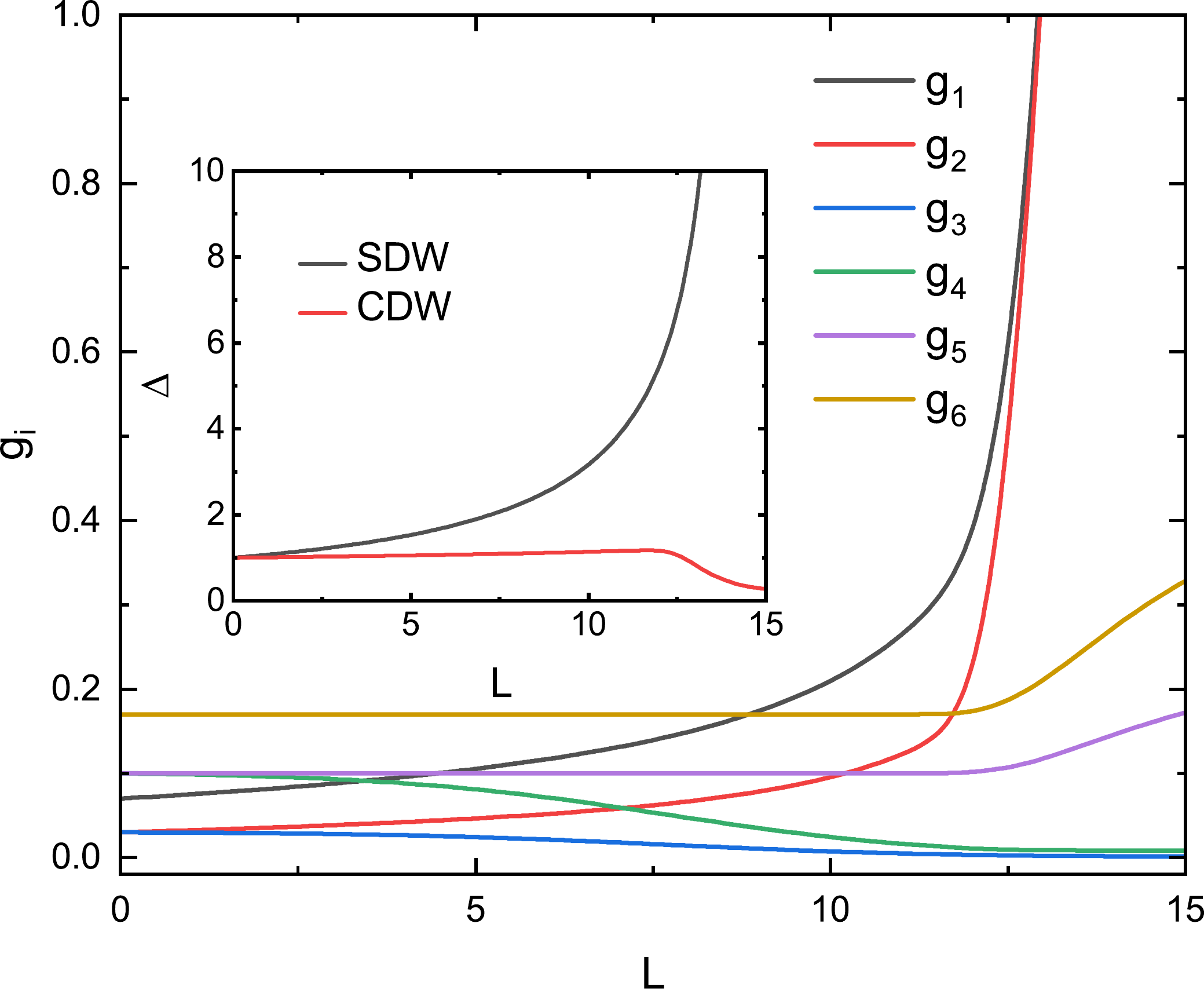} 
\includegraphics[width=0.35\textwidth]{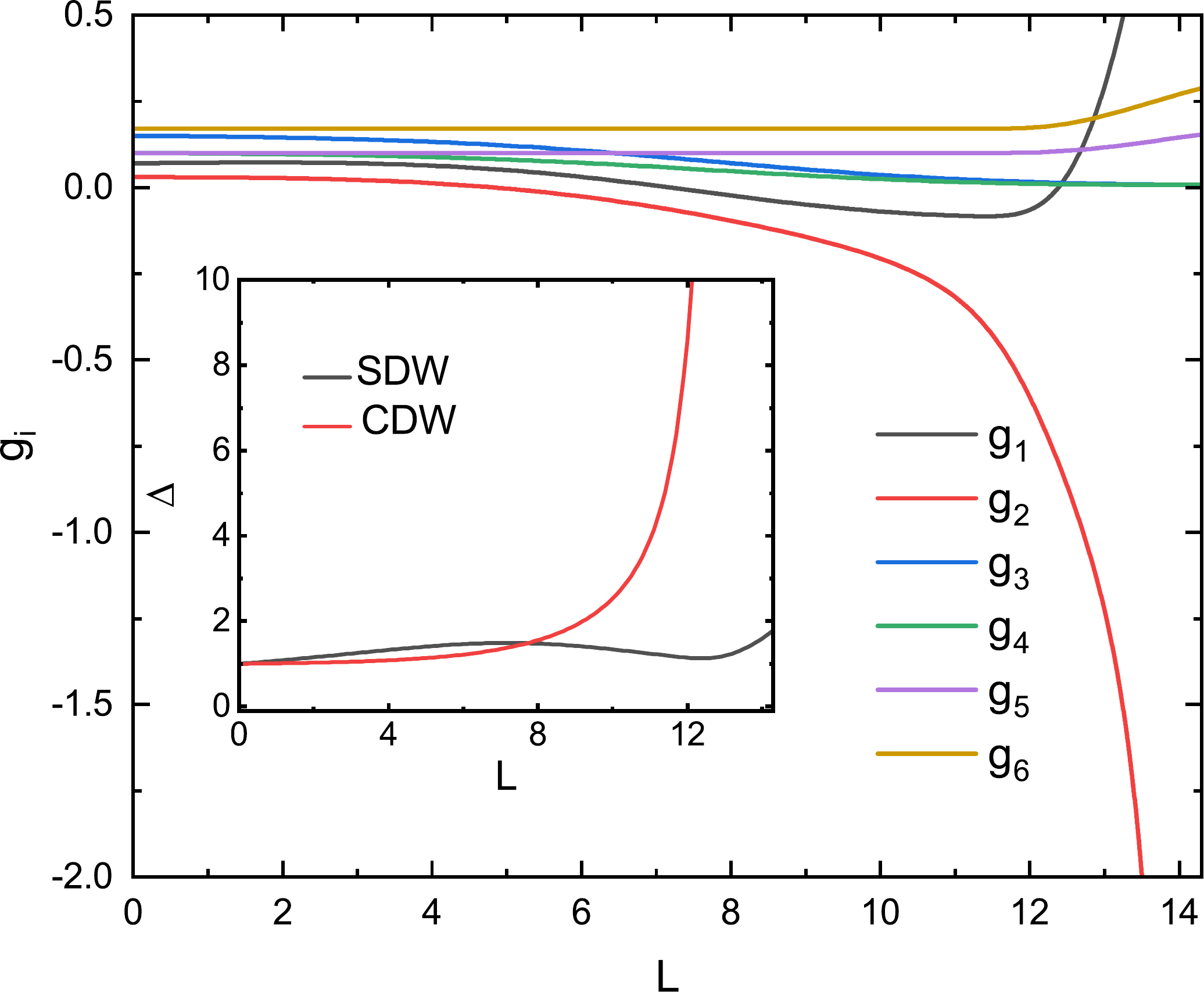}
\caption{For both figures  $g_1^0=0.07, g_2^0=0.03, g_4^0=0.1, g_5^0=0.1, g_6^0=0.17$, for the left figure $g_3^0=0.03$ while for the right figure  $g_3^0=0.15$. In both cases the T$_c$ increases by one order of magnitude. For the left figure the SDW T$_c =1.7 \times 10^{-6}$, while for the right one the CDW T$_c = 1.6 \times 10^{-6}$.}
\end{figure}
%

\end{widetext}

\begin{thebibliography}{99}
\bibitem{comment}
As the electron-electron interactions are short ranged and under the assumption that the Wannier functions are not too extensive, then the density-density interactions are expected to be of the same order as the exchange ones. In addition,  the coupling constants will  be significantly renormalized by two factors (i) when excited bands as well as the states of the lowest bands outside the patches are integrated out, and (ii) by electron-phonon coupling. It is reasonable to integrate out the phonons to obtain phonon-mediated electron-electron effective interaction, which renormalizes the values of coupling constants. As a result it is possible to obtain exchange interactions larger than density-density ones.
\end{thebibliography}
\end{document}